\documentclass[12pt]{article}
\usepackage{amsmath}
\usepackage{graphicx,psfrag,epsf}
\usepackage{enumerate}
\usepackage{natbib}
\usepackage{textcomp}
\usepackage[hyphens]{url} 
\usepackage{hyperref}
\providecommand{\tightlist}{%
  \setlength{\itemsep}{0pt}\setlength{\parskip}{0pt}}

\newcommand{\blind}{0}

\newcommand{\coursename}[1]{{\em Introduction to Data Science and Statistical Thinking}}

\newcommand{\school}[1]{Duke University}
\newcommand{\schoolshort}[1]{Duke}
\newcommand{\dsbox}[1]{\href{https://datasciencebox.org/}{datasciencebox.org}}
\newcommand{\github}[1]{\href{https://github.com/mine-cetinkaya-rundel/fresh-ds}{github.com/mine-cetinkaya-rundel/fresh-ds}}

\usepackage{xcolor}
\xdefinecolor{blue}{rgb}{0.28,0.65,0.9}
\hypersetup{
    colorlinks,
    linkcolor={blue},
    citecolor={blue},
    urlcolor={blue}
}
\usepackage{booktabs}
\usepackage{longtable}
\usepackage{array}
\usepackage{multirow}
\usepackage{wrapfig}
\usepackage{float}
\usepackage{colortbl}
\usepackage{pdflscape}
\usepackage{tabu}
\usepackage{threeparttable}
\usepackage{threeparttablex}
\usepackage[normalem]{ulem}
\usepackage{makecell}
\usepackage{xcolor}

\begin{document}

\def\spacingset#1{\renewcommand{\baselinestretch}%
{#1}\small\normalsize} \spacingset{1}


\if0\blind
{
  \title{\bf A fresh look at introductory data science}

  \author{
        Mine \c{C}etinkaya-Rundel \\
    School of Mathematics - University of Edinburgh,\\
Department of Statistical Science - Duke University, and RStudio\\
     and \\     Victoria Ellison \\
    Department of Statistical Science - Duke University\\
      }
  \maketitle
} \fi

\if1\blind
{
  \bigskip
  \bigskip
  \bigskip
  \begin{center}
    {\LARGE\bf A fresh look at introductory data science}
  \end{center}
  \medskip
} \fi

\bigskip
\begin{abstract}
The proliferation of vast quantities of available datasets that are
large and complex in nature has challenged universities to keep up with
the demand for graduates trained in both the statistical and the
computational set of skills required to effectively plan, acquire,
manage, analyze, and communicate the findings of such data. To keep up
with this demand, attracting students early on to data science as well
as providing them a solid foray into the field becomes increasingly
important. We present a case study of an introductory undergraduate
course in data science that is designed to address these needs. Offered
at Duke University, this course has no pre-requisites and serves a wide
audience of aspiring statistics and data science majors as well as
humanities, social sciences, and natural sciences students. We discuss
the unique set of challenges posed by offering such a course and in
light of these challenges, we present a detailed discussion into the
pedagogical design elements, content, structure, computational
infrastructure, and the assessment methodology of the course. We also
offer a repository containing all teaching materials that are
open-source, along with supplemental materials and the R code for
reproducing the figures found in the paper.
\end{abstract}

\noindent%
\vfill

\newpage
\spacingset{1.45} 

\hypertarget{introduction}{%
\section{Introduction}\label{introduction}}

\label{sec:intro}

How can we effectively and efficiently teach data science to students
with little to no background in computing and statistical thinking? How
can we equip them with the skills and tools for reasoning with various
types of data and leave them wanting to learn more? This paper describes
an introductory data science course that is our (working) answer to
these questions.

At its core, the course focuses on data acquisition and wrangling,
exploratory data analysis, data visualization, inference, modeling, and
effective communication of results. Time permitting, the course also
provides very brief forays into additional tools and concepts such as
interactive visualizations, text analysis, and Bayesian inference. A
heavy emphasis is placed on a consistent syntax (with tools from the
tidyverse), reproducibility (with R Markdown), and version control and
collaboration (with Git and GitHub). The course design builds on the
three key recommendations from \citet{nolan_lang2010}: (1) broaden
statistical computing to include emerging areas, (2) deepen
computational reasoning skills, and (3) combine computational topics
with data analysis. The goal of the course is to bring students from
zero experience to being able to complete a fully reproducible data
science project on a dataset of their choice and answer questions that
they care about within the span of a semester.

In Section \ref{sec:background} of this paper, we start with a review of
the most recent curriculum guidelines for undergraduate programs in data
science, statistics, and computer science. In this section we also
present a synopsis of the course content and structure of introductory
data science courses at four other institutions with the goal of
providing a snapshot of the current state of affairs in undergraduate
introductory data science curricula. In Section \ref{sec:course} we
outline the overall design goals of the \school{} introductory data
science course that is the focus of this article and discuss how this
course addresses current undergraduate curriculum guidelines in
statistics and data science. In Section \ref{sec:units} we expand on the
course content, flow, and pacing, and present examples of case studies
from the course. In Section \ref{sec:pedagogy} we detail the pedagogical
methods employed by this course, specifically addressing how these
methods can support a large class with students with a diverse range of
previous experiences in statistics and programming. Section
\ref{sec:computing} presents the computing infrastructure of the course,
Section \ref{sec:assessment} presents the methods of assessment, and
finally in Section \ref{sec:discussion} we provide a synthesis of where
this course sits in the landscape of introductory data science
curriculum guidelines, future design plans for the course, and
opportunities and challenges for faculty wanting to adopt this course.

\hypertarget{background-and-related-work}{%
\section{Background and related
work}\label{background-and-related-work}}

\label{sec:background}

An exact characterization of what the field of data science is meant to
encompass is still debated. However, in this paper we define data
science as the ``science of planning for, acquisition, management,
analysis of, and inference from data'' \citep{statsnsf2014}. We reviewed
four of the most recent curriculum guidelines for undergraduate programs
in data science, statistics, and computer science to assess how the case
study course ranks up against them.

While the 2013 Computer Science Curricula of the Association for
Computing Machinery (ACM) \citep{csguidelines2013} do not mention
suggestions for integrating data science into a computer science major,
the 2019 report by the ACM Task Force on Data Science Education
\citep{danyluk2019acm} gives suggestions of core competencies a
graduating data science student should leave with. Each competency
corresponds to one of nine data science knowledge areas: computing
fundamentals; data acquirement and governance; data management, storage,
and retrieval; data privacy, security, and integrity; machine learning;
big data; analysis and presentation; and professionalism. The report
also suggests that a full data science curriculum should integrate
courses in ``calculus, discrete structures, probability theory,
elementary statistics, advanced topics in statistics, and linear
algebra.'' We note, however, that this document was released as a draft
at the time of writing this manuscript.

Their recommendation for the first course is to introduce the
statistical analysis process starting with formulating good questions
and considering whether available data are appropriate for addressing
the problem, then conducting a reproducible data analysis, assessing the
analytic methods, drawing appropriate conclusions, and communicating
results. They also recommend that data science skills, such as managing
and wrangling data, algorithmic problem solving, working with
statistical analysis software, as well as high-level computing languages
and database management systems, be well integrated into the statistics
curriculum.

The 2016 Guidelines for Assessment and Instruction in Statistics
Education (GAISE) endorsed by the American Statistical Association also
does not make specific recommendations for introductory data science
courses, however the guidelines place emphasis on teaching statistics as
an ``investigative process of problem-solving and decision making'' as
well as giving students experience with ``multivariable thinking''
\citep{gaise2016}. The guidelines also recommend that students use
technology to explore concepts and analyze data, and suggest examples of
doing so using the R statistical programming language \citep{rproject}.

The Curriculum Guidelines for Undergraduate Programs in Data Science
suggest that the first introductory course for students majoring in data
science should introduce students to a high-level computing language
(they recommend R) to ``explore, visualize, and pose questions about the
data'' \citep{de2017curriculum}. Introduction to a high-level computing
language, data exploration and wrangling, basic programming and writing
functions, introduction to deterministic and stochastic modeling,
concepts of projects and code management, databases, and introduction to
data collection and statistical inference are among the suggested list
of topics for the first two courses in a data science major.
Furthermore, the guidelines propose that the introductory data science
courses be taught in a way that follows the full iterative data science
life cycle, ``from initial investigation and data acquisition to the
communication of final results.'' Finally, this report recommends ending
the course with a version-controlled, fully-reproducible, team-based
project, complete with a written and oral presentation. While the
\school{} course we describe in Sections \ref{sec:course} through
\ref{sec:discussion} was originally designed prior to the publication of
\citet{de2017curriculum}, the guidelines outlined in this report served
as inspiration for much of the updates to the course over the five years
that it has been taught.

In addition to curriculum guidelines, there exists a body of literature
on suggestions and case studies for integrating data science
computational skills into the general statistics curriculum.
\citet{nolan_lang2010} suggest including and discussing in detail
fundamentals in scientific computing with data, information
technologies, computational statistics (e.g., numerical algorithms) for
implementing statistical methods, advanced statistical computing, data
visualization, and integrated development environments into the
undergraduate statistics curriculum. \citet{hardin2015} and
\citet{baumer2015} provide case studies of data science courses that use
R as a computing language and have been implemented at various levels
within a statistics undergraduate major. \citet{dichev2017towards} and
\citet{brunner2016teaching} discuss single Python-based based
introductory data science case studies for courses without
prerequisites. \citet{dichev2016preparing} describe an introductory data
science course that teaches Python and R and that does not have any
prerequisites. Finally, while technically written for data science
graduate courses, \citet{hicks2018guide} promote teaching data science
via utilizing numerous case studies and emulating the process that data
scientists would use to answer research questions.

In their report titled ``Data Science for Undergraduates, Opportunities
and Options'', the National Academies of Sciences Engineering and
Medicine (NASEM) provide a wider survey of institutions that have
implemented stand-alone introductory data science courses designed to
serve as a general education requirement or garner general interest in
the field of data science \citep{NAP25104}. Three major challenges
identified in the report that are associated with teaching an
introductory data science course without any prerequisites are (1)
increasing student interest that is reflected in higher enrollment
numbers and the need to reconcile this with instructor availability, (2)
specific curriculum of the course varying from semester to semester
based on instructor expertise and interests, and (3) students with
diverse computing backgrounds thriving in a course with a
one-size-fits-all curriculum.

As part of our efforts to understand the landscape of undergraduate
introductory data science courses, we surveyed four courses that do not
require any student background in statistics or programming. These
courses are as follows:

\begin{enumerate}
\def\labelenumi{\arabic{enumi}.}
\tightlist
\item
  Foundations of Data Science (DATA 8) at University of California
  Berkeley
\item
  Foundations of Data Science at University of Cambridge
\item
  Introduction to Data Science (SDS 192) at Smith College
\item
  Data Science 101 (STATS 101) at Stanford University
\end{enumerate}

These courses were selected based on the ranking of the programs they
are taught in as well as the type of institution -- we wanted to capture
courses from a variety of institutions in terms of public/private,
US/non-US, research/liberal arts \citep{beststats2018, beststatsor2017}.
These were courses we were somewhat familiar with prior to data
collection and hence knew that they fit our requirements.

Table \ref{tab:class-components} gives a summary of the programming
languages used as well as a rough course content breakdown for these
four courses as well as the \school{} course that we discuss in further
details in the remainder of this manuscript.

\begin{table}

\caption{\label{tab:class-components}Summary of programming languages used in each course and the estimated breakdown of percent of class time spent on various course components.}
\centering
\begin{tabular}[t]{llllll}
\toprule
 & Duke & Berkeley & Cambridge & Smith & Stanford\\
\midrule
\rowcolor{gray!6}  Programming language & R & Python & Pseudocode & R, SQL & R\\
Data visualization & 15\% & 5\% & 0\% & 32\% & 10\%\\
\rowcolor{gray!6}  Data wrangling & 10\% & 15\% & 0\% & 36\% & 0\%\\
Other EDA & 10\% & 5\% & 0\% & 12\% & 10\%\\
\rowcolor{gray!6}  Inference & 20\% & 30\% & 25\% & 0\% & 50\%\\
\addlinespace
Modeling & 25\% & 20\% & 35\% & 0\% & 20\%\\
\rowcolor{gray!6}  Programming principles & 10\% & 10\% & 0\% & 5\% & 0\%\\
Mathematical foundations / theory & 5\% & 5\% & 35\% & 0\% & 0\%\\
\rowcolor{gray!6}  Communication & 5\% & 5\% & 0\% & 10\% & 10\%\\
Ethics & 0\% & 5\% & 5\% & 5\% & 0\%\\
\bottomrule
\end{tabular}
\end{table}

For each course, we surveyed the online course syllabus from a recent
semester and noted the lecture topic for each day of the course, the
programming language(s) used, and the assessed components. Then for each
course, we classified each day's lecture topic into one of nine content
categories given in Table \ref{tab:class-components}. Using these
classifications we calculated an approximate distribution of the amount
of lecture time spent on each of the nine content categories. Finally,
we contacted the instructors of these four courses and, based on their
feedback, adjusted our original content distribution estimates.

We first note that programming is a central role for each of these
courses. The courses at \school{}, Smith College, and Stanford
University teach R; and the course at UC Berkeley teaches Python. The
course at University of Cambridge is unique as it teaches only
pseudocode, although students are encouraged to learn Python on their
own time. In line with the greater focus that the Smith College course
places on data wrangling, SQL is also used in this course as well.

We allocated content in our rubric for ``Communication'' if the course
has a student project in which the students had to present their
findings. We note that the \school{}, Smith College, Stanford
University, and UC Berkeley courses all have some project presentation
element. No project component was mentioned for the University of
Cambridge course.

In addition, \school{}, Smith College, UC Berkeley, and University of
Cambridge courses all have some discussion on data ethics built into the
class.

We next note the differences in the extent to which each of these
courses make use of group assignments and assessments. At \school{}
students complete homework assignments and take-home exams individually,
and lab assignments and projects in groups. At Smith College students
work individually on homework assignments as well as on exams, they are
strongly encouraged students to work in pairs on the lab assignments,
and they work in groups for the projects. At Stanford University
students work individually on exams and homework assignments. At UC
Berkeley, the labs, homework assignments, and exams are completed
individually by the student, while the students are allowed to work with
one other student during the project. Finally, at University of
Cambridge, students take one exam that they complete individually.

We note the vast diversity of course content within each of these
classes compared to one another. For instance, Smith College emphasizes
the initial phases of the data science life cycle, such as data
visualization and data wrangling, whereas \school{}, UC Berkeley,
Stanford University, and University of Cambridge place more attention on
the middle phase of the data science life cycle, such as inference and
modeling. The University of Cambridge course places a heavier emphasis
on the mathematical foundations of data science than the other four
courses. Finally, while the \school{}, UC Berkeley, and University of
Cambridge courses place roughly equal focus on inference and modeling,
the Stanford University course places a much larger emphasis on
inference than on modeling.

Part of the reason for different levels of emphasis placed on different
phases of the data science life cycle that we observe among these
classes may be attributed to the differences in the primary audience the
course is designed for. For instance, \school{} course is designed to
provide a common (gateway) experience to students interested in the
Statistical Science major and minor or the interdisciplinary major in
Data Science. The Smith College course is a required course for
statistics majors while the UC Berkeley course is aimed at entry-level
students from all majors and the University of Cambridge course is
designed as a prerequisite for more advanced statistical and computer
science topics.

\hypertarget{the-course}{%
\section{The course}\label{the-course}}

\label{sec:course}

In this paper we describe an introductory data science course that is
designed to provide a common (gateway) experience to students interested
in the Statistical Science major and minor or the interdisciplinary
major in Data Science offered at \school{} called \coursename{}. A
version of this course has been offered as a seminar to first year
undergraduates each fall semester since the fall of 2014, with an
enrollment of 18 students at each offering under the title
\textit{Better Living with Data Science}. The course, with some
modifications for scale, was opened up to an audience of 80 students in
the Spring semester of 2018.

The main design goals were to create a course that is modern, that
places data front and center, that is quantitative without mathematical
prerequisites, that is different than high school statistics, and that
is challenging without being intimidating. The course emphasizes modern
and multivariate exploratory data analysis, and specifically data
visualization; starts at the beginning of the data analysis cycle with
data collection and cleaning; encourages and enforces thinking, coding,
writing, and presenting collaboratively; explicitly teaches best
practices and tools for reproducible computing; approaches statistics
from a model-based perspective, lessening the emphasis on statistical
significance testing; and underscores effective communication of
findings.

In addition, use of real data is a pillar of this course. Not only is
this strongly recommended in \citet{gaise2016}, but it also equips
students with the tools to answer questions of their own choosing as
part of their end-of-semester project.

\begin{figure}
\includegraphics[width=1\linewidth]{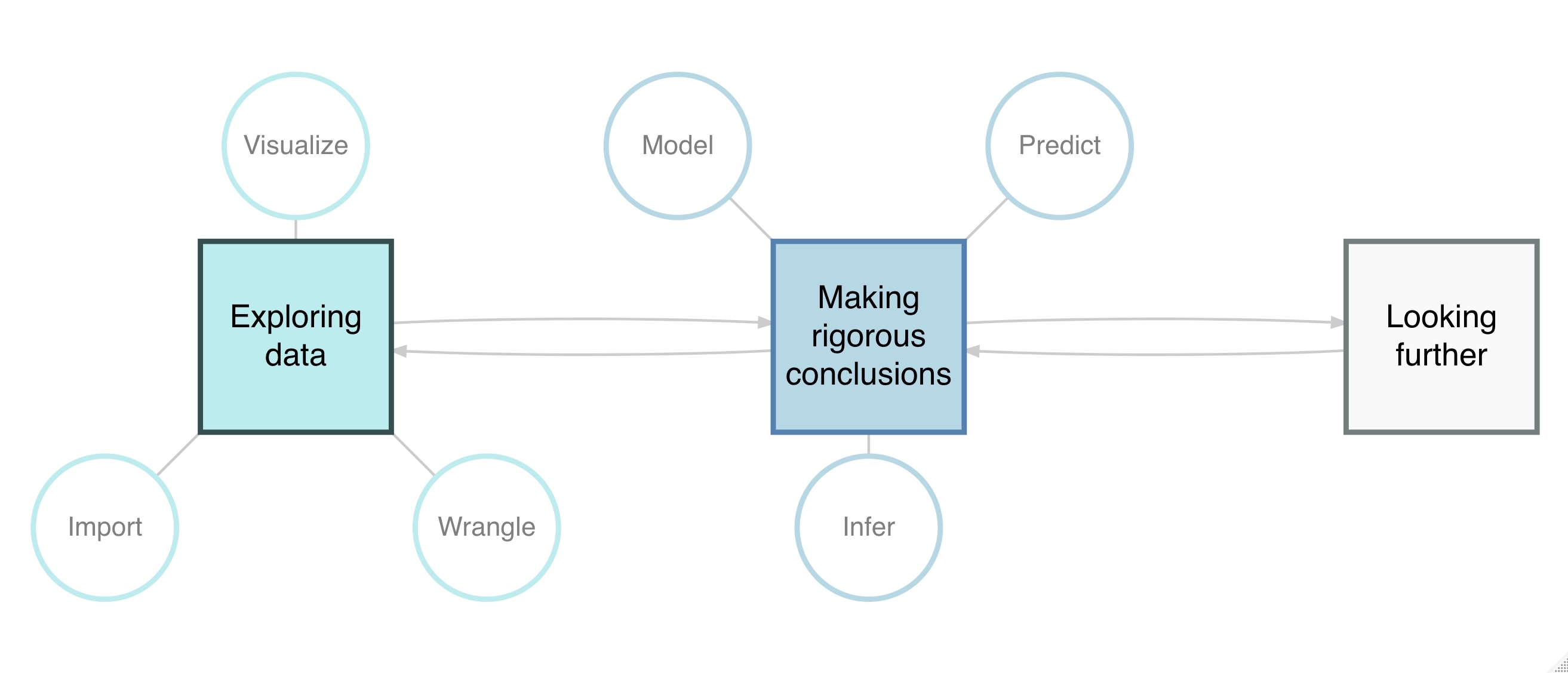} \caption{Flow of topics in \coursename{} at \school{}.}\label{fig:topicflow}
\end{figure}

Figure \ref{fig:topicflow} summarizes the flow of the three learning
units in STA 199: exploring data, making rigorous conclusions, and
looking forward. The arrows represent a continuous review and reuse of
previous material as new topics are introduced. The course ultimately
covers all steps of the full data science cycle presented in
\citet{wickham2016r}, which includes data import, tidying, exploration
(visualise, model, transform), and communication. In Section
\ref{sec:units} we describe in detail the topics covered in each of
these units.

\hypertarget{learning-units}{%
\section{Learning units}\label{learning-units}}

\label{sec:units}

The course is comprised of three learning units. The first two are
roughly of equal length, and the last one covers two weeks out of a
fifteen week semester.

\hypertarget{unit-1.-exploring-data}{%
\subsection{Unit 1. Exploring data}\label{unit-1.-exploring-data}}

This unit has three main foci: data visualization, data wrangling, and
data import.

The learning goals of the unit are as follows:

\begin{enumerate}
\def\labelenumi{\arabic{enumi}.}
\tightlist
\item
  Introduce the R statistical programming language via building simple
  data visualisations.
\item
  Build graphs displaying the relationship between multiple variables
  using data visualisation best practices.
\item
  Perform data wrangling, tidying, and visualisation using packages from
  the tidyverse.
\item
  Import data from various sources (e.g., CSV, Excel), including by
  scraping data off the web.
\item
  Create reproducible reports with R Markdown, version tracked with Git
  and hosted on GitHub.
\item
  Collaborate on assignments with team mates and resolve any merge
  conflicts that arise.
\end{enumerate}

On the first day of the course students log in to a web-based R session
and create a multivariate visualisation exploring how countries have
voted in the United Nations General Assembly on various issues such as
human rights, nuclear weapons, and the Palestinian conflict using data
from the \textbf{unvotes} package in R \citep{unvotes}. This is used as
an ice breaker activity to get students talking to each other about what
countries they are interested in exploring. The activity also gets them
creating and interpreting a data visualisation. Getting students to
create a data visualisation in R so quickly is made possible using
cloud-based computing infrastructure (which we describe in more detail
in Section \ref{sec:computing}) and a fully functional R Markdown
document. We call this the ``let them eat cake first'' approach, where
students first see an example of a complex data visualisation, which
they will be able to build by the end of this unit, and then slowly work
their way through the building blocks \citep{eat-cake}. This approach is
also presented in \citet{wang2017data}, which advocates for ``bringing
big ideas into intro stats early and often''.

There are two main reasons for starting data science instruction with
data visualisation. The first reason is that most students come in with
intuition for being able to interpret data visualizations without
needing much instruction. This means we can focus the majority of class
time initially on R syntax, and leave it up to the students to do the
interpretation. Later in the course, as students are getting more
comfortable with R and more advanced statistical techniques are
introduced, this scale tips where we spend more class time on concepts
and model interpretation and less on syntax. Second, it can be easier
for students to detect if they are making a mistake when building a
visualization, compared to data wrangling or statistical modeling.

In addition to the process of creating data visualisations, this unit
focuses on critiquing and improving data visualisations. After a brief
lecture on data visualisation best practices, that was designed in
collaboration with data visualisation experts at \school{}, we present
guidance for implementing these best practices in ggplot2 graphics. Each
team is given a flawed data visualisation as well as the raw data it is
based on. First, they critique the data visualisation and brainstorm
ways of improving it. Then, they (attempt to) implement their
suggestions for improvements. Finally, they present why and how they
improved their visualisations to the rest of the class. Since this
exercise happens early on in the semester, some teams fail to implement
all of their suggestions, but this ends up being a motivator for
learning. Additionally, multiple teams work on the same visualisation
and data, which makes the presentations valuable opportunities for
learning from each other. This exercise is described in further detail,
along with specific data sources and sample visualisations in
\citet{ccetinkaya2020drab}.

In the data wrangling and tidying part of Unit 1, we make heavy use of
the \textbf{dplyr} and \textbf{tidyr} packages for transforming and
summarising data frames, joining data from multiple data frames, and
reshaping data from wide to long / long to wide format. One example of a
data join is an exercise where country level data is joined with a
continent lookup table. This simple exercise presents an opportunity to
discuss data science ethics as some of the countries in the original
dataset do not appear in the continent lookup table (e.g., Hong Kong and
Myanmar) due to political reasons. The technical solution to this
problem is straightforward -- we can manually assign these countries to
a continent based on their geographic location. However we also discuss
that country-level datasets are inherently political as different
nations have different definitions of what constitutes a country -- an
example of how data processing workflow might be affected by data issues
\citep{NAP25104}. This data wrangling task is tied to a visualisation
exercise as well. By joining shapefile data to the country data we have,
we create choropleth maps as well. To simplify the exercise, we use the
\textbf{maps} package, along with ggplot2, for built-in shapefiles
instead of downloading these files from the web \citep{maps}.

Finally in Unit 1 we touch on data import. We start by introducing
commonly used data import options for reading rectangular data into R
(e.g., using \texttt{read\_csv()} or \texttt{read\_excel()} functions
from the \textbf{readr} and \textbf{readxl} packages). We then present
web scraping as a technique for harvesting data off the web using the
\textbf{rvest} package \citep{rvest}. We scrape data from
\href{https://www.opensecrets.org/}{OpenSecrets} (opensecrets.org), a
non-profit research group that tracks money in politics in the United
States. While the specific dataset we scrape changes from year to year,
the structure of the web scraping activity stays relatively constant:
first scrape data from a single page (containing data on a single voting
district, or single election year), convert the code developed for
scraping data from this single page into a function that takes a URL and
returns a structured data frame, and finally iterate over many similar
web pages (other voting districts, or other election years) using
mapping functions from the \textbf{purrr} package \citep{purrr}. We
usually end this exercise with a data visualisation created using the
scraped data that allows students to gain insights that would have been
impossible to uncover without getting the data off the web and into R.

In summary, this unit starts off with data visualisation on a dataset
that is already clean and tidy (and usually contained in an R package).
Then, we take one step back and learn about data wrangling and tidying.
Finally, we take one more step back and introduce both statistical and
computational aspects of data collection and reading data into R from
various sources.

\hypertarget{unit-2---making-rigorous-conclusions}{%
\subsection{Unit 2 - Making rigorous
conclusions}\label{unit-2---making-rigorous-conclusions}}

In Unit 1 students develop their skills for describing relationships
between variables, and the transition to Unit 2 is done via the desire
to quantify these relationships and to make predictions.

This unit is designed to achieve the following learning goals:

\begin{enumerate}
\def\labelenumi{\arabic{enumi}.}
\tightlist
\item
  Quantify and interpret relationships between multiple variables.
\item
  Predict numerical outcomes and evaluate model fit using graphical
  diagnostics.
\item
  Predict binary outcomes, identify decision errors and build basic
  intuition around loss functions.
\item
  Perform model building and feature evaluation, including stepwise
  model selection.
\item
  Evaluate the performance of models using cross-validation techniques.
\item
  Quantify uncertainty around estimates using bootstrapping techniques.
\end{enumerate}

We start off by introducing simple linear regression, but then quickly
move on to multiple linear regression with interaction effects since
students are already familiar with the idea that we need to examine
relationships between multiple variables at once to get a realistic
depiction of real world processes. We also introduce logistic
regression, albeit briefly. Prediction, model selection, and model
validation are introduced to pave the pathway for machine learning
concepts that students can dive further into in subsequent higher level
classes.

Finally in this unit we introduce the concept of quantifying
uncertainty, starting with uncertainty in slope estimates and model
predictions. We also touch on slightly more traditional introductory
statistics topics such as statistical inference for comparing means and
proportions. However, unlike many traditional introductory statistics
courses, inference focuses on confidence intervals, constructed using
bootstrapping only.

In designing this unit we had three goals in mind: (1) introduce models
with multiple predictors early, (2) touch on elementary machine learning
methods, and (3) de-emphasize the use of p-values for decision making.
The first goal addresses the 2016 GAISE recommendation for giving
students experience with multivariable thinking \citep{gaise2016}.
Additionally, introducing this topic early helps students frame their
project proposals (often due in the middle of this unit) by signalling
that this is a technique they might use in their projects. Teaching
logistic regression also proves to be invaluable in a course where
students later choose their own datasets and research questions for
their final projects. Each semester there are a considerable number of
teams who, as part of their project, want to tackle a task involving
predicting categorical outcomes, and familiarity with logistic
regression allows them to do so as long as they can dichotomize their
outcome. The second goal (touching on machine learning methods) presents
two opportunities. First, it enables a discussion on modeling binary
outcomes as both ``logistic regression'' (where we interpret model
output to evaluate relationships between variables) and ``binary
classification'' (where we care more about prediction than explanation).
Second, exposing students to foundational techniques like
classification, predictive modeling, cross-validation, etc. enables them
to start developing basic familiarity with machine learning approaches.
The third goal (de-emphasize the use of p-values for decision making) is
achieved by not covering null hypothesis significance testing in any
meaningful way. Traditional statistical inference topics are limited to
confidence intervals and decision errors that are presented in the
context of a logistic regression / classification. Students learn how to
construct confidence intervals using bootstrapping, and emphasis is
placed on interpreting these intervals in the context of the data and
the research question and we discuss decision making based on these
intervals. We also present decision making in the context of a
classification problem (a spam filter), where we explore the cost of
Type 1 and Type 2 errors to start building intuition around loss
functions.

One of the datasets featured in this unit comes from 18th century
auctions for paintings in Paris. In the case study of these paintings,
we explore relationships between metadata on paintings that were encoded
based on descriptions of paintings from over 3,000 printed auction
catalogues. These data include attributes like dimensions, material,
orientation, and shape of canvas, number of figures in the painting,
school of the painter, as well as whether the painting was auctioned as
part of a lot or on its own. The goal is to build a model predicting
price of paintings. However the data requires a fair amount of cleaning
before it can be used for building meaningful models. For example, some
of the categorical variables (e.g., material and shape of canvas) have
levels that are either misspelled or occur at low frequency. This offers
an opportunity for students to review data wrangling skills from the
previous unit while also learning about modeling. Additionally, the
response variable, price, is right skewed, which provides a nice
opportunity to introduce transformations. Finally, the dataset has over
60 variables, which means considering all interaction effects is not
trivial. Instead we explore interaction effects that the data experts
(art historians who created the dataset) have suggested. This provides
an opportunity for discussion around automated model selection methods
vs.~model building based on expert opinion.

Other datasets include professor evaluations and their ``beauty'' scores
(numerical, continuous outcome: evaluation score) and metadata on emails
(categorical, binary outcome: spam/not spam).

On the computational side, we use the \textbf{broom} package
\citep{broom} for tidy presentation of model output. Two features of
this package are especially well suited for the learning goals of this
course. First, regression output is returned as a data frame that makes
it easier to extract values from the output to include in reproducible
reports. This allows students to easily use inline R code chunks to
extract statistics like coefficient estimates or R-squared values from
model outputs and include in their interpretations, as opposed to
manually typing them out, which is recommended for reproducibility of
reports. Second, model summaries printed using the \texttt{tidy()}
function from the broom package do not contain the significant starts
that draw the attention to p-values. Note that it is possible to turn
these off in base R model summaries as well, but it is preferable to not
have them in the first place.

Like broom, other R packages introduced in this unit are part of the
\textbf{tidymodels} suite of packages, which is ``a collection of
packages for modeling and machine learning using tidyverse principles''
\citep{tidymodels}. These include \textbf{infer} for simulation-based
statistical inference and \textbf{modelr} for quantifying predictive
performance.

\hypertarget{unit-3---looking-forward}{%
\subsection{Unit 3 - Looking forward}\label{unit-3---looking-forward}}

This unit is designed to shrink or expand as needed depending on time
left in the semester. Each module is designed to cover one class period
and aims to provide a brief introduction to a topic students might
explore in higher level courses. One exception to this is an ethics
module, which kicks off the unit and is the only required component. In
this module we introduce ethical considerations around misrepresentation
in data visualizations and reporting of analysis results, p-hacking,
privacy, and algorithmic bias.

The remaining topics in the unit vary from semester to semester
depending on interests of the students and the instructor. In each class
period students are exposed to a few R packages that they use to engage
with specialised tasks (e.g., \textbf{flexdashboard} for building
dashboards \citep{flexdashboard}, \textbf{genius} for accessing song
lyrics \citep{genius}, \textbf{gutenbergr} for retrieving text from
books \citep{gutenbergr}, \textbf{shiny} for creating web apps
\citep{shiny}, \textbf{tidytext} for text analysis \citep{tidytext}).
Table \ref{tab:unit3-topics} lists topics covered in this unit in the
past, along with a brief synopsis.

\begin{table}

\caption{\label{tab:unit3-topics}Topics previously covered in Unit 3 of the course.}
\centering
\begin{tabular}[t]{>{\raggedright\arraybackslash}p{8em}>{\raggedright\arraybackslash}p{20em}>{\raggedright\arraybackslash}p{8em}}
\toprule
Topic & Synopsis & Duration\\
\midrule
\rowcolor{gray!6}  Data science ethics & Misrepresentation of results in data visualisations and reporting, data privacy and data breaches, gender bias in machine translated text, algorithmic bias and race in sentencing and parole length decisions. & 1-2 class periods\\
Interactive reporting and visualisation with Shiny & Introduce the basics of the shiny package for building interactive web applications and build a simple application for browsing data on movies. & 1 class period\\
\rowcolor{gray!6}  Building static dashboards & Build static dashboards using the flexdashboard package. & 1 class period\\
Building interactive dashboards & Build interactive dashboards using the shiny and flexdashboard packages. & 2 class periods\\
\rowcolor{gray!6}  Text mining & Perform basic text mining techniques (e.g., sentiment analysis, term frequency–inverse document frequency) using the tidytext package and data on song lyrics (retrieved with the genius package) or on books (retrieved with the gutenbergr package). & 1 class period\\
\addlinespace
Bayesian inference & Introduction to Bayesian inference as a way of decision making using data on sensitivity and specificity of breast cancer screening tests. & 1 class period\\
\bottomrule
\end{tabular}
\end{table}

\hypertarget{pedagogy}{%
\section{Pedagogy}\label{pedagogy}}

\label{sec:pedagogy}

In this section we discuss the various pedagogical choices (teamwork,
lectures sprinkled with hands-on exercises, computational labs, etc.) as
well as assessment components and feedback loops in the course. We
anticipate that instructors designing a similar course would be
especially interested in how we evaluate whether students in the course
achieve the outlined learning goals as well as a commentary on
assessment scalability for larger courses.

The pedagogical methods employed are tailored to several specific
aspects of the course. First, the course is relatively large in size
with about 80-90 students. Second, while the course has no statistical
or computing pre-requisites, students come into the course with very
diverse backgrounds -- some have no prior exposure to statistics or
computing while others may have already had a few classes in either of
the subjects, or both. As suggested by the literature
\citep{michaelsen2011team}, we employ several team-based learning
techniques to address the challenges of keeping a large lecture hall of
students with varying degrees of background knowledge both challenged
and engaged.

Within each lab section we aim to disperse students who have previously
learned some computing and/or statistics and those without any
background in these areas evenly amongst groups of four. In order to
gauge a student's prior background in statistics we have each student
complete a pretest before the course begins. We use the Comprehensive
Assessment of Outcomes in a First Statistics course (CAOS) test, an
online test developed by Assessment Resource Tools for Improving
Statistics Thinking (ARTIST) project
\href{https://app.gen.umn.edu/artist/}{app.gen.umn.edu/artist} intended
to assess students on the key concepts that any student coming out of an
introductory statistics course would be expected to know. We use a
combination of scores from this test as well as information on computing
experience to roughly classify students into three categories of ``has
background'', ``doesn't have any background'', and ``somewhere in
between''. We then assign one student who is identified as ``has
background'', one who is identified as ``doesn't have any background'',
and two students from the ``somewhere in between'' categories to teach
team. In choosing which students to pick from these categories to place
into each team, we take into account self-reported information collected
via a ``Getting to Know You'' survey, such as interests, (planned)
major, personal pronouns, etc. We aim to create demographically diverse
teams where each student shares some attributes with at least one other
student in the team. The team assignment process is carried out
manually, which presents challenges as the class size grows. However
since students stay in these teams throughout the entire semester,
taking extra care during the team formation process is a worthwhile
investment for reducing team dynamic issues that might arise later in
the semester.

The method of content delivery is mostly lecture, and student feedback
on whether they desire more or less content to be delivered during the
actual lecture has been mixed. Future iterations of this course may seek
to decrease the amount of new content delivered to the students during
the lecture and shift the students first exposure to the material to
pre-class assignments or videos. This shift is informed by the body of
literature which suggests better learning and better student
satisfaction in introductory statistics courses taught using a flipped
classroom approach where students completed relatively simple reading
and answered reading quiz questions prior to class and completed
hands-on exercises in class
\citep{wilson2013flipped, winquist2014flipped}. In place of new content
delivered in lecture, future iterations of the course may incorporate
more extensive group application exercises into the class time, allowing
students to get individual feedback on their current understanding from
their peers, the TAs, and the instructor.

\hypertarget{computing-and-infrastructure}{%
\section{Computing and
infrastructure}\label{computing-and-infrastructure}}

\label{sec:computing}

In this section we discuss the computing choices made in the course,
including infrastructure, syntax, and tools. In this section we will
detail the computing infrastructure used in the course (access to
RStudio in the cloud) and provide pedagogical justifications for the
decisions made in setting up this infrastructure. Additionally, we will
provide a road map of the computational toolkit, outlining when and why
students get introduced to each new package or software.

\hypertarget{rstudio-cloud}{%
\subsection{Seamless onboarding with RStudio
Cloud}\label{rstudio-cloud}}

This course follows the recommendations outlined in
\citet{cetinkaya_rundel} for setting up a computational infrastructure
to allow for pedagogical innovations while keeping student frustration
to a minimum.

The most common hurdle for getting students started with computation is
the very first step: installation and configuration. Regardless of how
well detailed and documented instructions may be, there will always be
some difficulty at this stage due to differences in operating system,
software version(s), and configurations among students' computers. It is
entirely possible that an entire class period can be lost to
troubleshooting individual student's laptops. An important goal of this
class is to get students to create a data visualization in R within the
first ten minutes of the first class. Local installation can be
difficult to manage, both for the student and the instructor, and can
shift the focus away from data science learning at the beginning of the
course.

Access to R is provided via RStudio, an integrated development
environment (IDE) that includes a viewable environment, a file browser,
data viewer, and a plotting pane, which makes it less intimidating than
the bare R shell. Additionally, since it is a fully fledged IDE, it also
features integrated help, syntax highlighting, and context-aware tab
completion, which are all powerful tools that help flatten the learning
curve.

Rather than locally installing R and RStudio, students in this course
access RStudio in the cloud via \href{https://rstudio.cloud/}{RStudio
Cloud} (rstudio.cloud), a managed cloud instance of the RStudio IDE. The
main reason for this choice is reducing friction at first exposure to R
that we described above.

When you create an account on RStudio Cloud you get a workspace of your
own, and the projects you create here are public to RStudio Cloud
members. You can also add a new workspace and control its permissions,
and the projects you create here can be public or private. A natural way
to set up a course in RStudio Cloud is using a private workspace. In
this structure, a classroom maps to a workspace. Once a workspace is set
up, instructors can invite students to the workspace via an invite link.
Workspaces allow for various permission levels which can be assigned to
students, teaching assistants, and instructors. Then, each
assignment/project in the course maps to an RStudio Cloud project.

Another major advantage of this setup over local installation of R and
RStudio is that workspaces can be configured to always use particular
versions of R and RStudio as well as a set of packages (and particular
versions of those packages). This means the computing environment for
the students can easily be configured by the instructor, and always
matches that of the instructor, further reducing frustration that can be
caused by instances of the student running the exact same code as the
professor but getting errors or different results.

\hypertarget{literate-programming-and-reproducibility-with-r-markdown}{%
\subsection{Literate programming and reproducibility with R
Markdown}\label{literate-programming-and-reproducibility-with-r-markdown}}

Building on \emph{literate programming} \citep{Knuth1984}, R Markdown
provides an easy-to-use authoring framework for combining statistical
computing and written analysis in one computational document that
includes the narrative, code, and the output of an analysis
\citep{xie2018r}. On the first day of the course, upon accessing the
computing infrastructure via RStudio Cloud as described in Section
\ref{rstudio-cloud}, students are presented with a fully functional R
Markdown document including a brief but not-so-simple data analysis that
they can knit to produce an in-depth data visualization. Then, by
updating just one parameter in the R Markdown document, they can produce
a new report with a new data visualization. This process of an early win
is made possible with R Markdown in a way that would be much harder to
accomplish typing code in the console or even with the use of a
reproducible R script. We are able to introduce students to R Markdown
before any formal R instruction thanks to the very lightweight syntax of
the markdown language, and by providing a fully functional document that
is guaranteed to knit and display results for each student regardless of
their personal computing setup.

Throughout the course students use a single R Markdown document to
write, execute, and save code, as well as to generate data analysis
reports that can be shared with their peers (for teamwork) or
instructors (for assessment). Early on in the course we facilitate this
experience by providing them templates that they can use as starting
points for their assignments. Throughout the semester this scaffolding
is phased out, and the final project assignment comes with a bare-bones
template with just some suggested section headings.

The primary benefit of using R Markdown in statistics and data science
instruction are outlined in \citet{Baumer2014} as restoring the logical
connection between statistical computing and statistical analysis that
was broken by the copy-and-paste paradigm. Use of this tool keeps code,
output, and narrative all in one document, and in fact, makes them
inseparable.

\hypertarget{clean-and-consistent-grammar-with-the-tidyverse}{%
\subsection{Clean and consistent grammar with the
tidyverse}\label{clean-and-consistent-grammar-with-the-tidyverse}}

The curriculum makes opinionated choices when it comes to specific
programming paradigms introduced to students. Students learn R with the
\textbf{tidyverse}, an opinionated collection of R packages designed for
data science that share an ``underlying design philosophy, grammar, and
data structures'' \citep{wickham2019welcome}. The most important reason
for this choice is the cohesiveness of the tidyverse packages. The
expectation is that learning one package makes it easier to use the
other due to these shared principles. Tidyverse code is not necessarily
concise, but the course aims to teach students to maximize readability
and extensibility of their code instead of minimizing the number of
lines to accomplish a task.

\hypertarget{version-control-and-collaboration-with-git-and-github}{%
\subsection{Version control and collaboration with Git and
GitHub}\label{version-control-and-collaboration-with-git-and-github}}

One of the learning goals of this course is that how you got to a data
analysis result is just as important as the result itself. Another goal
is to give students exposure to and experience using software tools for
modern data science. Use of literate programming with R Markdown gets us
part of the way there, but implicit in the idea of reproducibility is
collaboration. The code you produce is documentation of the process and
it is critical to share it (even if only with yourself in the future).
This is best accomplished with a distributed version control system like
Git \citep{bryan2018excuse}. In addition, Git is a widely used tool in
industry for code sharing. According to an industry-wide Kaggle survey
of data scientists conducted by Kaggle, 58.4\% of over 6,000 respondents
said Git was the main tool used for sharing code in their workplace
\citep{kaggle_2017}.

In this class we have adopted a top down approach to teaching Git --
students are \emph{required} to use it for \emph{all} assignments.
Additionally, GitHub is used as the learning management system for
distributing and collecting assignments as repositories. Based on best
practices outlined in \citet{cetinkaya_rundel}, we structure the class
as a GitHub organization, and a starter private repository is created
per student/team per assignment, and we use the \textbf{ghclass} package
for instructor management of student repositories \citep{ghclass}.

Students interact with Git via RStudio's project based Git GUI. We teach
a simple centralized Git workflow which only requires the student to
know how to perform simple actions like \texttt{push}, \texttt{pull},
\texttt{add}, \texttt{rm}, \texttt{commit}, \texttt{status}, and
\texttt{clone}. Focusing on this core functionality helps flatten the
learning curve associated with a sophisticated version control tool like
Git for students who are new to programming
\citep{fiksel2019using, beckman2020implementing}. Early on in the
course, we also engineer situations in which students encounter problems
while they are in the classroom so that the professor and teaching
assistants are present to troubleshoot and walk them through the process
in person.

We note that GitHub can also be used as an early diagnostic tool to
identify students that may struggle in the course later on. We pulled
the data on all commits made by students in the Spring 2018 cohort of
the course. The usage of these data was given an exemption from IRB
review by \school{} Campus Institutional Review Board.

\begin{figure}
\includegraphics[width=1\linewidth]{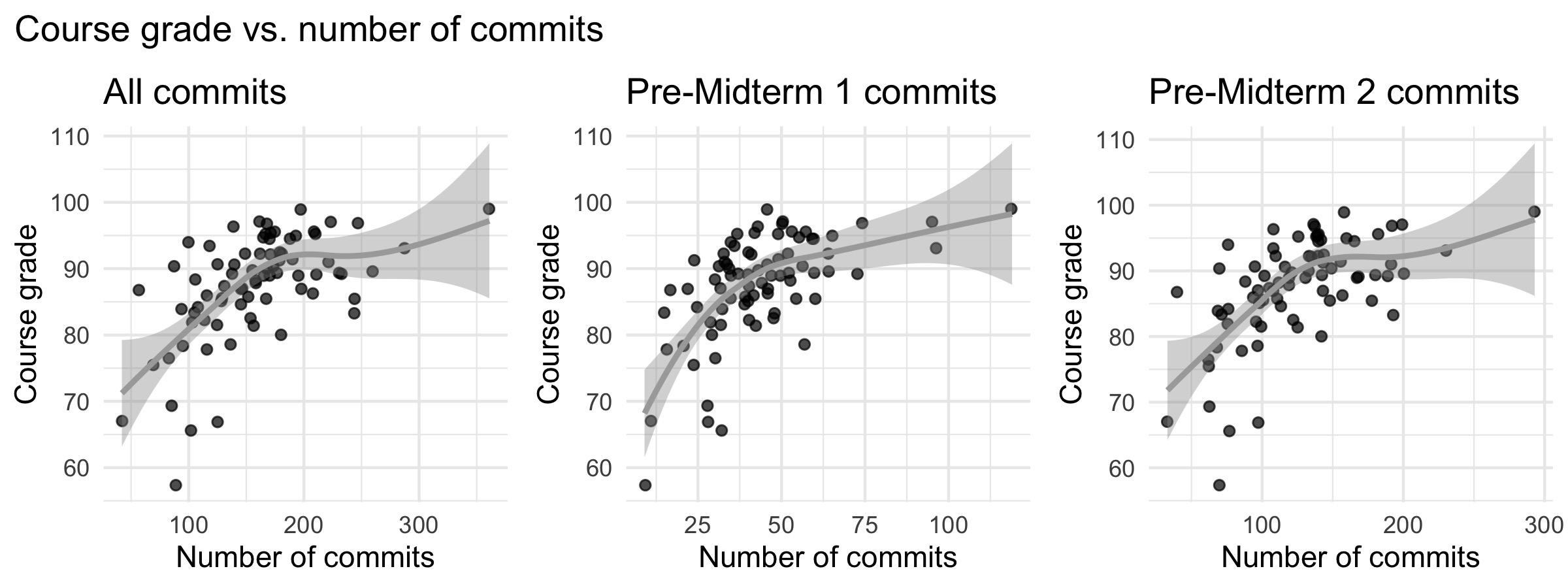} \caption{Relationship between number of commits and final course grade for each student at three time points in the semester.}\label{fig:commit_scatter}
\end{figure}

Figure \ref{fig:commit_scatter} displays three plots created with these
data. The plot on the left shows the relationship between number of
commits made by each student throughout the entire semester and their
final course grade (out of 100 points). The plot in the middle and on
the right also display the final course grade on the y-axis but the
number of commits made by each student are calculated at earlier time
points in the semester (before the first midterm for the plot in the
middle, and before the second midterm for the plot in the right). We can
see a positive relationship in each of the plots, levelling off at 100
points (since it is not possible to score higher than 100 points in the
course). While number of commits, alone, should not be considered an
indication of course performance, these plots suggest that one can
identify students with low numbers of commits as those who will
potentially not perform well in the course, and reach out to them early
on and offer support and help.

Incorporation of version control and collaboration with Git and GitHub
into the introductory data science classroom not only benefits students
by teaching them skills desired by potential employers, but it also cuts
down on the administrative work required to distribute, grade, and
return assignments, which can now be spent providing in-depth feedback,
working with students, and updating course material.

\hypertarget{assessment}{%
\section{Assessment}\label{assessment}}

\label{sec:assessment}

This course uses five methods of assessment, each designed with the
incoming student with no background in statistics or computing in mind.
First, we have weekly computing labs which are completed in groups. With
these labs, students without any coding background can benefit from the
prior coding experience of other students in the group. However, in an
effort to make sure that each student, including those with no computing
experience, has weekly practice in coding we also assign individual
homework assignments as well. Finally, because programming plays a
central role in the course, we incorporate coding exercises into the
midterm exams. In order to accommodate first-time programmers in which a
timed coding exam may prove to be infeasible, the midterms are set as
take-home exams and the students are allowed to use books, notes, and
the internet to complete them.

Participation also factors into the final grade of students in the
course. In addition, voluntary participation such as answering a
question or being called on to answer a question has been shown to cause
higher anxiety in large introductory courses than working in groups on
in-class exercises \citep{england2017student}. Therefore, instead of
relying solely on a potentially subjective measure of voluntary
participation, participation scores of students in this class are made
up of a check / no check type grade on their team-based in-class
application exercises (they get a check if they were in class for the
day) as well as their engagement on the online course discussion.

Many of the assignments and assessments in the course are designed to
prepare students for the final project, which, in a nutshell, asks
students to ``Pick a dataset, any dataset, and do something with it.''
The actual assignment, of course, goes into a lot more details than
this, but ultimately students are asked to work in teams to pick a new
(to them) dataset and an accompanying research question and answer the
question using methods and tools they learned in the course. We
specifically ask them to not feel pressured to apply everything they
learned, but to be selective about which method(s) they use. They are
also encouraged to try methods, models, and approaches that go beyond
what they learned in the course and additional support for implementing
these is provided during office hours.

There are three main reasons for assigning this team-based final
project. First, in a class where students start off with no prerequisite
knowledge, it is hugely rewarding for them to see that they can go from
zero to full fledged collaborative and reproducible data analysis within
the span of a semester, and hopefully this leaves them wanting to learn
more. Second, for the most part, teamwork results in a better final
product than students would accomplish individually. And lastly, teams
are more adventurous than individual students, and are more likely to
venture outside of what they learned in the class and learn new tools
and methods to complete their projects.

Teams turn in a project proposal roughly one-month before the final
project is due with their data and proposed analysis. These proposals
are reviewed carefully and feedback is provided to the students. Teams
can choose to revise their proposals based on the feedback, and thereby
increase their score on the proposal stage of the project. The final
deliverables of the project are a 10-minute presentation during the
scheduled final exam time and a write-up that goes into further depth
than the presentation can in the allotted time. The final write-up is an
R Markdown file, but unlike the earlier assignments, code chunks are
turned off so that only the prose and the output/plots are visible to
the reader. This encourages students to pay attention to wording,
grammar, and most importantly flow since their narrative isn't
interrupted with large chunks of code.

\hypertarget{discussion}{%
\section{Discussion}\label{discussion}}

\label{sec:discussion}

The impact of this course at \school{} has been profound. Increasing
numbers of students coming out of this course continuing their studies
in statistics after this course helped provide impetus to update and
modernize the computational aspects of the second statistics course in
regression. For example, the regression course now also uses the
tidyverse syntax, students complete assignments using R Markdown, and
use version control with Git, and collaborate and submit assignments on
GitHub. Additionally, the course has served as a way to start building
bridges between the introductory statistical science and computer
science curricula, accelerating the formation of an interdepartmental
major in data science, where students are provided an option to build a
full undergraduate curriculum in data science but mixing and matching
from a list of prescribed courses from the two departments. In addition
to students wanting to pursue a degree in statistics and/or data
science, this course also serves a large number of students from the
social and natural sciences as well as the humanities. The course now
satisfies the introductory statistics requirement of many majors (e.g.,
political science, public policy, economics), and hence we expect to see
trickle down effects of starting with data science within the
statistical and computational learning goals of these disciplines as
well.

As \citet{baumer2015} put it so well, ``{[}i{]}f data science represents
the new reality for data analysis, then there is a real risk to the
field of statistics if we fail to embrace it.'' Statistics departments
are at a huge advantage for offering courses that can prepare students
to embrace and extract meaning from modern data: we have faculty
proficient in statistical inference, modeling, and computing.
Traditionally these three pillars of statistics came together in higher
level courses, but we believe that it's time to flip things around.
Offering an introductory course like the one described in this article
can introduce students to data science early on, as early as their first
semester in college due to not having any prerequisites for the course.
This will not only help drum up interest in the topic (and hence in
statistics) but also provide a pathway for students to start interacting
meaningfully with data and developing their computational skills while
concurrently taking mathematical prerequisites needed for a statistics
major, such as calculus, linear algebra, etc.

It has been ten years since \citet{nolan_lang2010} suggested that
``{[}i{]}t is our responsibility, as statistics educators, to ensure our
students have the computational understanding, skills, and confidence
needed to actively and wholeheartedly participate in the computational
arena.'' \emph{Introduction to Data Science and Statistical Thinking} is
designed to address this goal early on, and to introduce students to
statistical thinking through computing with data. While this course
alone is not sufficient to equip students with all of the computing
skills \citet{nolan_lang2010} outlines, it serves as a solid foundation
to build on.

One of the biggest challenges in designing this course has been deciding
which topics to include, especially in the second unit on making
rigorous conclusions. Some topics that are commonly covered in
introductory statistics courses are intentionally left out in order to
make room for increased emphasis on computing and computational
workflows. For example, this course places less emphasis on null
hypothesis significance testing and the Central Limit Theorem compared
to a traditional introductory statistics course. While we touch on
p-values as one way of making decisions based on statistical
information, we don't demonstrate how to calculate them in various
settings. Similarly, the Central Limit Theorem is only referenced in
relation to some of the common characteristics of bootstrap
distributions. So far, we only have anecdotal evidence that students who
take a course on regression after completing the introductory data
science course about their experience in the regression course. The
evidence suggests that they have sufficient statistical background to
succeed in the regression course and do not appear to be less prepared
than their peers who completed a traditional introductory statistics
course. Future research could help inform the downstream effects of
introduction to the discipline of statistics via this course and how
student learning outcomes in the statistics major compare to other
starting points.

In designing the course we had one more ambition: to make all course
materials openly licensed and freely available to the statistics and
data science instructor community. All course content (lecture slides,
homework assignments, computing labs, application exercises, and sample
exams) as well as materials on pedagogy and infrastructure setup to help
instructors who want to teach this curriculum can be found at \dsbox{}.

Beyond the challenges that come with designing any new course, there are
a few aspects of this course that we believe might present challenges
for instructors who want to adopt this course. First, while the
foundational skills in data science are well established, the technical
and implementation details, such as which R package should you use, can
be a moving target. Staying current with these active developments is
rewarding, but can be time consuming.

Second, teaching this curriculum involves engaging with technical
logistics that may be outside of the comfort zone of many instructors.
Much of this is addressed by professionally managed, web-based services
(e.g., RStudio Cloud) as well as tooling developed specifically to help
manage course logistics (e.g., the ghclass package). A willingness to
tackle unexpected technical difficulties (e.g., a student getting stuck
on an undecipherable Git error) using a combination of Googling and
copying and pasting from Stack Overflow will help. One can view this as
an opportunity as well -- live debugging sessions where an instructor
models how they search for answers on the web can be valuable learning
experiences for students.

Finally, the topics presented in this course are substantially different
than those in a traditional introductory statistics or introductory
probability course. This course provides less exposure to mathematical
statistics topics (e.g., the Central Limit Theorem, distributions,
probability) in favour of computational data analysis skills. As such,
it is important that the second course in a program is updated to
accommodate students coming in with different backgrounds, which will
require buy in from departmental faculty. We strongly believe that
statistics and data science programs that leverage and reinforce these
skills throughout the rest of the curriculum will ultimately produce
stronger graduates.

\hypertarget{supplementary-materials}{%
\section{Supplementary materials}\label{supplementary-materials}}

\label{sec:supplemental}

Supplemental materials for the article, including details on the data
collection process and the R code for reproducing the figures found in
the paper, can be found on GitHub at \github{}.

\hypertarget{acknowledgements}{%
\section*{Acknowledgements}\label{acknowledgements}}
\addcontentsline{toc}{section}{Acknowledgements}

We thank the editors, the associate editors, and anonymous reviewers for
their helpful comments and suggestions. We would also like to thank Ben
Baumer from Smith College, John Christopher Duchi from Stanford
University, David Wagner from University of California Berkeley, and
Damon Wischik from University of Cambridge for providing information on
their introductory data science courses.

\bibliographystyle{agsm}
\bibliography{fresh-ds.bib}

\end{document}